\documentclass[12pt]{article}
\usepackage{latexsym,epsfig,graphics}
\newcommand{\be}{\begin{equation}}
\newcommand{\ee}{\end{equation}}
\newcommand{\bq}{\begin{eqnarray}}
\newcommand{\eq}{\end{eqnarray}}
\begin{document}
\begin{titlepage}
\today          \hfill 
\begin{center}

\vskip .5in

{\large \bf A New World Sheet Field Theory}
\footnote{This work was supported 
 by the Director, Office of Science,
 Office of High Energy  Physics, 
 of the U.S. Department of Energy under Contract 
DE-AC02-05CH11231.}
\vskip .50in


\vskip .5in
Korkut Bardakci\footnote{Email: kbardakci@lbl.gov}

{\em Department of Physics\\
University of California at Berkeley\\
   and\\
 Theoretical Physics Group\\
    Lawrence Berkeley National Laboratory\\
      University of California\\
    Berkeley, California 94720}
\end{center}

\vskip .5in

\begin{abstract}

A second quantized field theory on the world sheet is developed
for summing planar graphs of the $\phi^{3}$ theory. This is in contrast
to the earlier work, which was based on first quantization. The ground
state of the model is investigated with the help of a variational ansatz.
 In complete agreement with standard perturbation theory,
the infinities encountered in carrying out this calculation can be
eliminated by the renormalization of the parameters of the model.
 We also find that,
as in the earlier work, in the ground state, graphs form a dense network
(condensate) on the world sheet. 

\end{abstract}
\end{titlepage}

\newpage
\renewcommand{\thepage}{\arabic{page}}
\setcounter{page}{1}

\noindent{\bf 1. Introduction}

\vskip 9pt

 This paper is the continuation of a series of papers
devoted to the summation of field theory graphs on the world sheet [1,2,3,4],
but at the same time, it is a fresh start.
As in the previous work, the starting point is 't Hooft's seminal paper [5],
 which puts planar Feynman graphs of the $\phi^{3}$ field theory on the world
sheet parametrized by the light cone variables. The world sheet is again
discretized , and
 extensive use is made of the
 fermions introduced in  [6] to ennumerate
graphs. But other than these  features, the present article differs
substantially from the earlier work on the same problem.

All  the previous papers on this subject,
 starting with [1],  used the formalism of first quantization:
The momenta flowing through the graph were treated as fields and
quantized. This approach was inspired by the well known
quantization of free strings on the world sheet, and in fact, the
idea behind the two parallel treatments was to build a bridge between
field theoery and string theory. In contrast, the present article is
based on a second quantized Fock space approach, very much along the
lines suggested in 't Hooft's original paper [5]. As explained in
section 2, on the world sheet,
a typical graph of the $\phi^{3}$ theory can be pictured as a collection
of parallel solid lines, which form the boundaries of the
 propagators (see Fig.1). The idea is to define fields $\phi$ and
$\phi^{\dagger}$ (eq.(6)), which annihilate and create these lines,
and construct the corresponding Fock space. This is analogous to the
usual second quantization of field theory, where the corresponding
field operators create and annihilate particles. Here, the analogue of
particles are the points on the solid lines  that mark
the boundaries of the propagators. From this perspective, $\phi^{\dagger}$
and $\phi$ create and annihilate boundaries.

The Fock space construction outlined above has one serious defect: It
has infinitely many unwanted states. They correspond to multiple
solid lines at the same location, and also to unallowed configuration of
propagators. In section 3, 
making crucial use of the world sheet fermions [6], we show how to overcome
 this problem. Consequently,
in addition to the $\phi$'s, the Fock space must also include the
fermions. In sections 3 and 4, we construct the free and the
interacting Hamiltonians in terms of these operators. The corresponding time
development operator is then shown to generate all all the
planar graphs of $\phi^{3}$ on the world sheet.

At this point, one can pose a natural question: What is the motivation
for inventing a new formalism for summing planar graphs? We list a couple  
responses to this question:\\
a) It is always a good idea to to try to find  many different
approaches to a difficult problem, with their different advantages and
disadvantages.\\
b) In our opinion, the new formalism is both simpler than the previous
one, and also it is better founded. The difficulties encountered in the
old formalism were discussed extensively in reference [4]; we will not repeat
this discussion here except to note that they all arose from technical issues
connected with first quantization. In contrast, the new approach based on
second quantization is both simple and encounters no such problems.
As an added bonus, in contrast to the old approach, it is easy
to introduce a finite mass.\\
c) In section 5, we study the ground state of the system, using a
variational approximation scheme. In this calculation, we encounter
ultraviolet divergences, which have to be renormalized. In the critical
dimension of space-time, which is six for the $\phi^{3}$ theory, the
divergences are the standard ones familiar from perturbation theory;
namely, logarithmic divergence in the coupling constant and quadratic
divergence in the mass. The model can be renormalized in the usual
fashion by absorbing these divergences into the bare parameters.
Also, the well known asymptotic freedom of the theory
comes out of the variational calculation. This is all very satisfactory,
especially compared to the earlier work, where renormalization of the model
was always a problem.

Of course, the idea behind the whole program is not just to reproduce
some results of perturbation theory but to go beyond them. The variational
calculation was mainly set up to compute the 
parameter $\rho_{0}$, defined by eqs.(5) and (26). This parameter
measures the density of the solid lines (boundaries) on the world sheet, 
which will be defined precisely at the beginning of section 5. Any finite
order graph consists of a finite number of solid lines, which form a set
of zero density on the world sheet. Therefore, the corresponding $\rho_{0}$ 
vanishes. A non-zero value for this parameter means that the world sheet is
densely occupied by the graphs, and the contribution of high (infinite)
order graphs dominate the ground state. Another and perhaps more familiar
way is to think of $\rho_{0}$ as an order parameter for phase transition.
A non-vanishing value for it signals the formation of a new
non-perturbative  phase corresponding to a densely occupied world sheet.
 In fact, the variational
calculation carried out in section 5 gives a non-zero value for $\rho_{0}$,
showing that the model is in this new phase.

In ending this section, we would like to stress that the reformulation
of planar $\phi^{3}$ as a world sheet 
field theory was the crucial first step.
Without such a formulation, we would not know how to define the concept
of the density of Feynman graphs, let alone introduce an order parameter
for a new phase. The simple variational ansatz of eq.(27) also relies
heavily on the world sheet picture.

\vskip 9pt

\noindent{\bf 2. A Brief Review}
\vskip 9pt

The generic\footnote{ By generic, we mean a graph where the external
momenta do not have any symmetry pattern. For a discussion, see [4].} 
   planar graphs of $\phi^{3}$ theory
 in the mixed light cone representation of 't Hooft [5] have a particularly
simple form, which we briefly review here. The world sheet is parametrized
by the two coordinates
$$
\tau=x^{+}=(x^{0}+x^{1})/\sqrt{2},\,\,\,\sigma=p^{+}=(p^{0}+p^{1})/\sqrt{2}.
$$
A general planar graph is represented by a collection of horizontal solid
lines (Fig.1), where the n'th solid line carries a $D$ dimensional
transverse momentum ${\bf q}_{n}$.
\begin{figure}[t]
\centerline{\epsfig{file=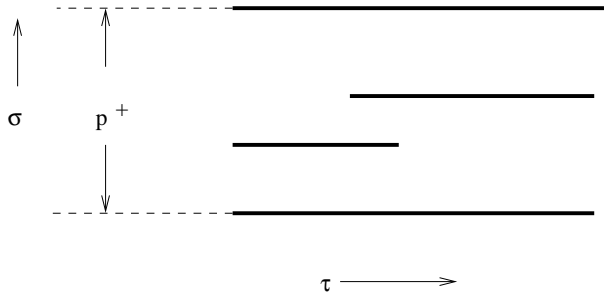,width=8cm}}
\caption{A Typical Graph}
\end{figure}
 Two adjacent lines labeled by n and
n+1 represent the light cone propagator
\be
\Delta(p_{n})=\frac{\theta(\tau)}{2 p^{+}}\,\exp\left(-i\tau\,\frac{
{\bf p}_{n}^{2}+m^{2}}{2 p^{+}}\right),
\ee
where ${\bf p}_{n}={\bf q}_{n}-{\bf q}_{n+1}$. A factor of $g$, the coupling
constant, is inserted at the beginning and at the end of each line,
where the interaction takes place.

For various technical reasons, it is very convenient to discretize the
coordinate $\sigma$ in steps of length $a$, which is equivalent to
compactifying the light cone coordinate $x^{-}=(x^{0}-x^{1})/\sqrt{2}$
at radius $R=1/a$. This type of compactification was first introduced
in connection with the M theory [7,8]; it can be viewed as an infinite boost
of the more standard compactification of a spacelike direction. In this
 paper, we will exclusively work with the discretized world sheet;
 on the other hand, the time coordinate $\tau$
will remain continuous.

We also have to specify the boundary conditions to be imposed on the
world sheet. For simplicity, the coordinate $\sigma$ is compactified
by imposing periodic boundary conditions at $\sigma=0$ and $\sigma=p^{+}$,
where $p^{+}$ is the total $+$ component of the momentum flowing
through the whole graph. In contrast, the boundary conditions at
$\tau=\pm \infty$ will be left free; in the Hamiltonian approach that
we are going to adopt, they will not play any role.

There is a useful way of visualizing the discretized world sheet,
pictured in Fig.2.
\begin{figure}[t]
\centerline{\epsfig{file=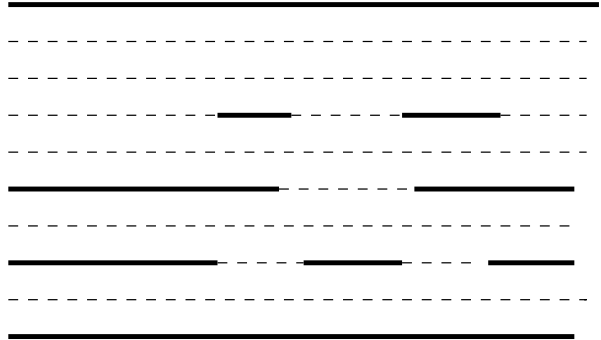,width=8cm}}
\caption{Solid And Dotted Lines}
\end{figure}
 In this picture, the world sheet consists of
horizontal dotted and solid lines, spaced a distance $a$ apart. The
boundaries of the propagators are marked by solid lines, same as in
Fig.1, and the bulk is filled with dotted lines. Ultimately, one has to 
integrate over all possible locations and lengths of solid lines, as well as 
over the transverse momenta they carry. The integrations over the locations
of the solid lines can be conveniently incorporated into the action or the
Hamiltonian formalism by introducing a two component fermion field
$\psi_{i}(\sigma,\tau), i=1,2$, and its adjoint $\bar{\psi}_{i}$ [6].
These fields satisfy the usual anticommutation relations
\be
\left[\psi_{i}(\sigma,\tau),\bar{\psi}_{i'}(\sigma',\tau)\right]_{+}=
\delta_{i,i'}\,\delta_{\sigma,\sigma'}.
\ee
The fermion with $i=1$ lives on the dotted lines and the one with $i=2$
lives on the solid lines. On an uninterrupted line, they satisfy the time
evolution equation
\be
\partial_{\tau}\psi_{i}=0,\,\,\partial_{\tau}\bar{\psi}_{i}=0.
\ee
On the other hand,
when the interaction takes place, there is a transition from $i=1$ to
$i=2$ or vice versa. The operators that generate these transitions are
given by\footnote{Here we are using the letter $\sigma$ for both the world
sheet coordinate and also for Pauli matrices. Hopefully, there should be
no confusion about the dual use of this letter.}
\be
\rho_{\pm}=\bar{\psi}\sigma_{\pm}\psi,
\ee
where
$$
\sigma_{\pm}=(\sigma_{1}\pm \sigma_{2})/2.
$$
For future use, we also note that, the operator
\be
\rho=\frac{1}{2}\,\bar{\psi}(1-\sigma_{3})\psi,
\ee
is equal to one on the solid lines and zero on the dotted lines.

\vskip 9 pt

\noindent{\bf 3. The Free Hamiltonian}

\vskip 9pt

We start by defining the Hilbert space, 
which should completely specify the $\sigma$ cross section of
 any given graph at a fixed value of the time
$\tau$ . At each site, labeled by a discrete value of $\sigma$,
there is either a dotted or a solid line, and a single fermion is placed
at that site with $i=1$ (spin up)
 in case of a dotted line, and with $i=2$ (spin down) in case
of a solid line. For a solid line, one has also to specify the
momentum ${\bf q}$ flowing through it, and for this purpose, we
introduce the  bosonic field $\phi(\sigma,\tau,{\bf q})$, and its conjugate
$\phi^{\dagger}(\sigma,\tau,{\bf q})$, which
at time $\tau$ respectively annihilate
and create a solid line carrying momentum ${\bf q}$ 
located at site labeled by $\sigma$. They satisfy
  the commutation relations
\be
\left[\phi(\sigma,\tau,{\bf q}),\phi^{\dagger}(\sigma',\tau,{\bf q'})\right]
=\delta_{\sigma,\sigma'}\,\delta({\bf q}-{\bf q'}).
\ee

We now have both the fermionic and bosonic operators needed to construct
the Fock space, and it remains to define the vacuum state. This corresponds
 to a state with only dotted lines (an empty world sheet), and it satisfies,
\be
\rho_{+}(\sigma)|0\rangle =0,\,\,\,\phi(\sigma,{\bf q})|0\rangle =0.
\ee
These conditions are initially imposed at a fixed time $\tau$
 for all values of $\sigma$ and ${\bf q}$. In equations of this type,
we usually do not explicitly write the dependence on time. Of course,
the equations of motion ensure that these conditions are satisfied for
all times.

It is now easy to construct the Fock space in the standard way. For example,
the state corresponding to two solid lines
 (a propagator) located at $\sigma_{1}$ and $\sigma_{2}$, carrying momenta
${\bf q}_{1}$ and ${\bf q}_{2}$, is given by,
\be
|1,2\rangle =\left(\phi^{\dagger}(\sigma_{1},{\bf q}_{1})\,
\rho_{-}(\sigma_{1})\right)
\left(\phi^{\dagger}(\sigma_{2},{\bf q}_{2})
\,\rho_{-}(\sigma_{2})\right)|0\rangle.
\ee
Notice that, to preserve the relation between the solid and the dotted
lines and the fermion spin (index i), the field $\phi$ is always
accompanied by the field $\rho_{+}$ at the same point, and the field
$\phi^{\dagger}$ by $\rho_{-}$ at the same point.

At this point, it might appear that the introduction of the fermions
was redundant. After all, we could have tried to construct the Fock
space using only the operators $\phi$ and $\phi^{\dagger}$. However,
the construction without the fermions runs into a serious problem.
In the Hilbert space without the fermions, there will be states
corresponding to multiple solid lines at the same site, created by the
repeated application of $\phi^{\dagger}$ at the same $\sigma$.
For example, the state
$$
\prod_{k=1}^{k=n}\phi^{\dagger}(\sigma,{\bf q}_{k})|0\rangle
$$
corresponds an n-tuple solid line at site $\sigma$, which has no
world sheet counterpart. This multiple  line clearly carries
a momentum
$$
\sum_{k=1}^{k=n} {\bf q}_{k},
$$
but since already there is a single  line carrying the same
momentum, the existence of multiple solid lines leads to an infinite
overcounting of the states.  At this point, the fermions come to the
rescue. Because of Fermi statistics
\be
\left(\rho_{-}(\sigma)\right)^{n}=0
\ee
for $n \geq 2$. Since we associate each $\phi^{\dagger}$ with a factor
of $\rho_{-}$ at the same site, states corresponding to multiple lines
automatically vanish.

The absence of multiple lines (solid or dotted) can be written in the form
of an equation of constraint to be imposed on the Hilbert space. We define
the  operator
\be
n(\sigma,\tau)=\int d{\bf q}\,
\phi^{\dagger}(\sigma,\tau,{\bf q})\,\phi(\sigma,\tau,{\bf q}),
\ee
which counts the number of solid lines located at $\sigma$. We also recall
 that
the operator $\rho(\sigma,\tau)$, defined earlier (eq.(5)), is one if there is
a solid line at $\sigma$, zero otherwise. It follows that, imposing the
constraint
\be
\rho(\sigma,\tau)-n(\sigma,\tau)=0,
\ee
 eliminates multiple lines. This constraint is first imposed at a fixed 
$\tau$ for all $\sigma$, and in order to extend it to  all $\tau$, we have
to show that it commutes with the Hamiltonian. We will do so once we have
the Hamiltonian at hand.

With these preliminaries out of the way, we turn to the construction of
the Hamiltonian for the free theory, where both the solid and dotted lines are
eternal, and the graphs are just a collection of free propagators. Consider
the state corresponding to a single propagator defined by eq.(8). If we
tentatively set,
\be
H_{0}=\frac{1}{2} \sum_{\sigma'>\sigma} \int d{\bf q} \int d{\bf q'}\,
\frac{({\bf q}-{\bf q'})^{2}+ m^{2}}
{\sigma'-\sigma}\,\phi^{\dagger}(\sigma,
{\bf q})\,\phi(\sigma,{\bf q})\,\phi^{\dagger}(\sigma',{\bf q'})\,
\phi(\sigma',{\bf q'}),
\ee
it easy to show that
\be
H_{0}|1,2\rangle =\int d{\bf q}_{1} \int d{\bf q}_{2}\, 
\frac{({\bf q}_{1}-{\bf q}_{2})^{2}+ m^{2}}{2\,|\sigma_{1}-\sigma_{2}|}.
\ee
 The time evolution operator
$$
\exp\left(- i \tau H_{0}\right)
$$
then reproduces the free propagator of eq.(1), except for the prefactor
$1/(2 p^{+})$. This missing factor will later be included in the
interaction Hamiltonian.

Although our proposal for the free Hamiltonian gives the right answer
for a single propagator, it does not work in the case of more than one
propagator. For example, consider two propagators, whose boundaries are
located at $\sigma_{1}$, $\sigma_{2}$ and $\sigma_{3}$, as shown in
Fig.3 .
\begin{figure}[t]
\centerline{\epsfig{file=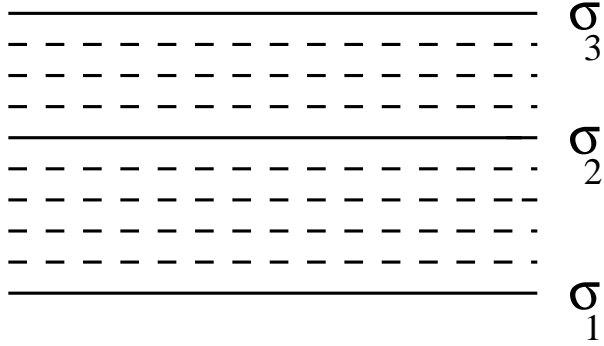,width=8cm}}
\caption{Two Propagators}
\end{figure}
 If we apply the free Hamiltonian given by eq.(12) to the state
corresponding to this figure, in addition to the two propagators with
momenta ${\bf q}_{1}-{\bf q}_{2}$ and ${\bf q}_{2}-{\bf q}_{3}$, we also
get a third unwanted propagator with momentum ${\bf q}_{1}-{\bf q}_{3}$.
The rule is that the propagators are associated with adjacent solid
lines such as $1,2$ and $2,3$, and not with non-adjacent ones, such as
$1,3$. To implement this rule automatically in the expression for $H_{0}$,
we define, for any two lines, solid or dotted, located at $\sigma_{i}$
and $\sigma_{j}$, with $\sigma_{j}>\sigma_{i}$,
\be
\mathcal{E}(\sigma_{i},\sigma_{j})=\prod_{k=i+1}^{k=j-1}
\left(1- \rho(\sigma_{k})\right),
\ee
where $\rho$ is given by eq.(5). If $\sigma_{j}<\sigma_{i}$, $\mathcal{E}$
is defined to be zero\footnote{ A similar function
 was introduced for the same reason in [4].}. From its definition,
$\mathcal{E}(\sigma_{i},\sigma_{j})$ is equal to one only if two solid lines
are located at $\sigma_{i}$ and $\sigma_{j}$, seperated by only dotted
lines, as in Fig.4.
\begin{figure}[t]
\centerline{\epsfig{file=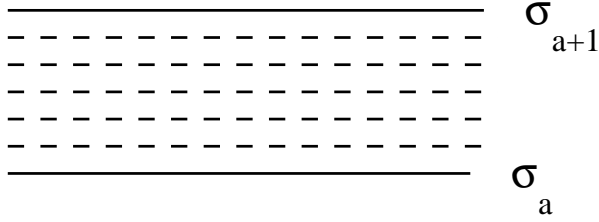,width=8cm}}
\caption{Solid Lines Seperated By Only Dotted Lines}
\end{figure}
 Otherwise, it is zero. In the example discussed above, 
$\mathcal{E}(\sigma_{i},\sigma_{j})$ is one for the combinations 
$i=1, j=2$ and $i=2, j=3$, and it vanishes for the unwanted combination
$i=1, j=3$. With the help of this operator, we write the correct version
of eq.(12):
\bq
H_{0}&=&\frac{1}{2}\sum_{\sigma,\sigma'}\int d{\bf q} \int d{\bf q'}\,
\frac{\mathcal{E}(\sigma,\sigma')}{\sigma'-\sigma} 
\left(({\bf q}-{\bf q'})^{2}+ m^{2}\right)\nonumber\\
&\times&\phi^{\dagger}(\sigma,
{\bf q})\,\phi(\sigma,{\bf q})\,\phi^{\dagger}(\sigma',{\bf q'})\,
\phi(\sigma',{\bf q'}).
\eq
By applying $H_{0}$ to a state with several solid lines,
one can easily show that there is no
longer the problem of unwanted propagators. We note the fermions
play a dual role: In addition to preventing the formation of
multiple solid lines, they eliminate the unwanted configurations of
the propagators.

Finally, by making use of the canonical algebra (eq.(6)), and 
 the commutation relation
$$
\left[\rho(\sigma,\tau),\rho(\sigma',\tau)\right]=0,
$$
one can verify that
\be
\left[(\rho- n),H_{0}\right]=0.
\ee
 This shows that the constraint imposed on the states by
eq.(11) is consistent with the time evolution.

\vskip 9pt

\noindent{\bf 4. The Interaction}

\vskip 9pt

The interaction takes place at the point of transition between a solid and
a dotted line. These transitions can be generated by  the
interaction Hamiltonian composed of two terms,
\be
H_{I}=g\,\sum_{\sigma}\int d{\bf q}\left(\rho_{+}(\sigma)\,\phi(\sigma,
{\bf q}) +\phi^{\dagger}(\sigma,{\bf q})\, \rho_{-}(\sigma)\right),
\ee
one of which converts a dotted line into a solid one and the other
does the opposite. Also
notice that, in accordance with the rule stated
after eq.(8), $\phi^{\dagger}$ is accompanied by a factor of $\rho_{-}$ and
$\phi$ by $\rho_{+}$. Then, using the commutation relations
$$
\left[\rho(\sigma),\rho_{\pm}(\sigma')\right]=\mp\,\delta_{\sigma,\sigma'\,}
\rho_{\pm}(\sigma),
$$
it is easy to verify that
\be
\left[(\rho- n),H_{I}\right]=0.
\ee

 This takes care of the interaction, but as
we have explained earlier, there are still the missing $1/(2 p^{+})$
prefactors of the propagators that have to be incorporated
into the interaction vertices. Consider the two types of vertices,
corresponding to the beginning and ending of a solid line, pictured in
Fig.5. The solid lines are labeled as 1, 2 and 3, and the momenta
that enter the vertex are labeled by the corresponding pair of indices
12, 23, and 13 respectively. As explained in reference [4], attaching
a factor of
\be
V=\frac{1}{\sqrt{8\,p_{12}^{+}\,p_{23}^{+}\,p_{13}^{+}}}=
\frac{1}{\sqrt{8(\sigma_{2}-\sigma_{1})(\sigma_{3}-\sigma_{2})
(\sigma_{3}-\sigma_{1})}}
\ee
to each vertex, takes care of the missing prefactors.

We face, however, the following problem: How do we attach $V$ to
$H_{I}$? It is clear that we should identify $\sigma_{2}$ with
the interaction point $\sigma$ in eq.(17), but what about $\sigma_{1,3}$?
 We have to remember that $H_{I}$
acts on a state with solid lines  located at some sites labeled, say,
by $\sigma_{a}$. The sites labeled by
 $\sigma_{1}$ and $\sigma_{3}$ in Fig.5a must then coincide
with the two adjacent solid lines labeled by $\sigma_{a}$ and $\sigma_{a+1}$
for some $a$.
\begin{figure}[t]
\centerline{\epsfig{file=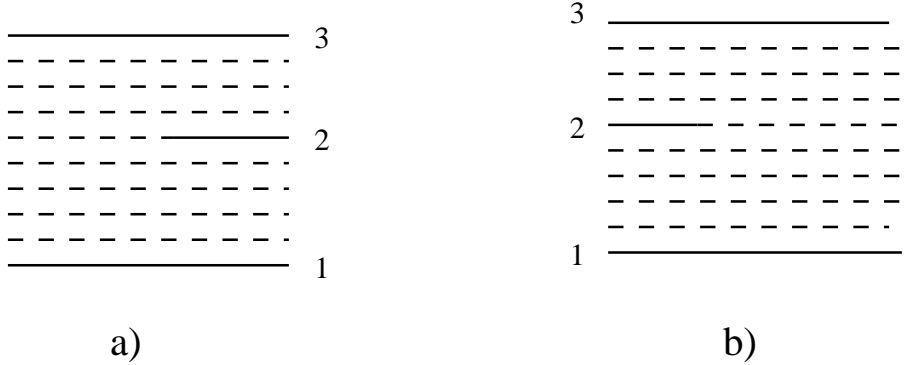,width=12cm}}
\caption{The Two $\phi^{3}$ Vertices}
\end{figure}
 To enforce this constraint, we attach a projection operator
$\mathcal{F}(\sigma_{1},\sigma_{3})$ to $V$. This operator is one
when $\sigma_{1}$ and $\sigma_{3}$ are located on two adjacent solid lines
of the state, that is, when they are seperated by only dotted lines.
 Otherwise, it vanishes. We already faced a
 similar problem in constructing the free Hamiltonian, and we define
a projection operator closely related to $\mathcal{E}$ (eq.(14)):\footnote{
The presence of two extra factors of $\rho$ in the definition of 
$\mathcal{F}$, or rather the absence of these
factors in $\mathcal{E}$, deserves an explanation. As explained above,
these factors are needed in the vertex to ensure that the corresponding sites
are occupied by solid lines. In contrast, in eq.(15), there is no need to
introduce these factors, since the combination
of fields $\phi^{\dagger}\phi$  at $\sigma$ and $\sigma'$
 guarantees that the lines located at these sites are solid.}
\be
\mathcal{F}(\sigma_{1},\sigma_{3})=\rho(\sigma_{1})\,\mathcal{E}(\sigma_{1},
\sigma_{3})\,\rho(\sigma_{3}),
\ee
and with the help of $\mathcal{F}$, we construct the vertex
\be
\mathcal{V}(\sigma_{2})=\sum_{\sigma_{1}<\sigma_{2}}\sum_{\sigma_{3}>
\sigma_{2}}\frac{\mathcal{F}(\sigma_{1},\sigma_{3})}{\sqrt{8\,
(\sigma_{2}-\sigma_{1})\,(\sigma_{3}-\sigma_{2})\,(\sigma_{3}-\sigma_{1})}}.
\ee
It is then easy to see that the projection operator lines up $\sigma_{1}$
and $\sigma_{3}$ correctly with the adjacent solid lines of the state on
which it acts. This vertex has to be attached to $H_{I}$ of eq.(17) to supply
the missing $1/(2 p^{+})$ factors. With this modification, the final form
of the interaction Hamiltonian is,
\be
H_{I}=g\sum_{\sigma}\int d{\bf q}\left(\rho_{+}(\sigma)\,
\phi(\sigma,{\bf q})\,\mathcal{V}(\sigma)+ \mathcal{V}(\sigma)\,
\phi^{\dagger}(\sigma,{\bf q})\,\rho_{-}(\sigma)\right),
\ee
and the full Hamiltonian is given by,
\be
H= H_{0}+ H_{I}.
\ee
The above Hamiltonian, combined with the constraint (11), generates all the
planar graphs on the world sheet.

Although we will stick with the Hamiltonian approach in this paper, if so
desired, one can easily switch to the path integral approach based on the
action
\be
S=\int d\tau\,\left(\sum_{\sigma}\left(i\bar{\psi}\partial_{\tau} \psi+
i \int d{\bf q}\,\phi^{\dagger} \partial_{\tau}\phi\right) - H(\tau)\right).
\ee
The constraint (11) has then to be imposed on the states at some initial time.

\vskip 9pt

\noindent{\bf 5. The Variational  Approximation}

\vskip 9pt

The Hamiltonian developed in the last section is exact but  it is also
quite complicated. Clearly, to make progress, one has to resort to some
approximation scheme. What we are
trying to calculate is the structure of the ground state of the system,
and for this purpose,
 the variational approach is the natural approximation scheme,
although it is also possible to do an equivalent mean field calculation.
In particular, as we have explained in the introduction, we are
interested in phenomenon of the condensation of graphs on the world
sheet, by which we mean the following: Consider the limit of a large
but finite number of lines, that is
\be
N_{0}= p^{+}/a
\ee
is large but finite. Here, $p^{+}$ total $+$ momentum flowing through
the graph and $a$ is the spacing between the lines. If in this limit the
 solid lines constitute a finite fraction of the total number of lines,
we say that a condensate of solid lines has formed. Another equivalent
way of defining the condensate is to use $\rho$ (eq.(5)) as the order
parameter. Consider the ground state expectation value of $\rho$,
\be
\rho_{0}=\langle \rho(\sigma,\tau)\rangle.
\ee
We will assume the ground state is invariant under the translations
of $\sigma$ and $\tau$,  so that $\rho_{0}$ does not depend on these
variables. By its definition, $\rho_{0}$ measures the quantity we are
interested in, namely, the probability of an average line in the ground
state of the system being solid. In any fixed order of perturbation
expansion, and in the limit of large $N_{0}$, $\rho_{0}$ is effectively
zero, since the vast majority of the world sheet lines will be dotted. 
On the other hand, a finite $\rho_{0}$ means that the ground state is
dominated by graphs whose order grow with $N_{0}$. We will find that,
within our approximation scheme, $\rho$ acquires a finite expectation
value, signaling the formation of the condensate. In carrying out this
calculation, we will also learn how to renormalize the model within
this approximation, and we will compare it to the standard results from
perturbation theory.

The ansatz for the ground state that will be used in the variational
calculation is,
\be
|s_{v}\rangle=\prod_{\sigma}\left(\int d{\bf q}\,A({\bf q})\,
\phi^{\dagger}(\sigma,{\bf q})\,\rho_{-}(\sigma) + B\right)|0\rangle,
\ee
where the product extends over all the sites and the vacuum state on the
right is defined by (7). It easy to see that since $A$ and $B$ are $\sigma$
independent, $|s_{v}\rangle$ is the direct product of identical states
located at each site. It is therefore just about the simplest non-trivial
ansatz for the ground state one can think of. $A$ and $B$ satisfy the
normalization condition
$$
\int d{\bf q}\,|A({\bf q})|^{2}+ |B|^{2}=1.
$$
In fact, each term in the above equation has a probability interpretation.
From the definition of $\rho_{0}$, it follows that
\be
\int d{\bf q}\,|A({\bf q})|^{2}=\rho_{0}
\ee
is the probability of having a solid line at any site, and
\be
|B|^{2}=1-\rho_{0}
\ee
is the probability of having a dotted line. The variational ansatz therefore
represents a world sheet uniformly (independent of $\sigma$) populated by
an arbitrary superposition of solid and dotted lines.

Following the standard variational principle, we will try to minimize
the expectation value
$$
\langle s_{v}|H|s_{v}\rangle
$$
by setting its variation with respect to $A({\bf q})$ and $B$ equal to
zero. We will take both $A$ and $B$ to be real, which can be done without
any loss of generality, and set
$$
B=\sqrt{1- \rho_{0}}.
$$
Also, assuming that the ground state is rotationally invariant, $A$ can only
 depend  on the length of the vector ${\bf q}$.
 The normalization condition will be taken care of
by defining
\be
I(A,\rho_{0},\lambda)= \langle s_{v}|H|s_{v}\rangle
+ \lambda \left(\int d{\bf q}\,A^{2}({\bf q})-\rho_{0}\right),
\ee
and varying $I$ instead of $\langle s_{v}|H|s_{v}\rangle$.
Here $\lambda$ is a Lagrange multiplier.

Next, we need the expectation value of
$$
H= H_{0}+H_{I}
$$
 in the variational state. After a straightforward calculation, we have, 
\be
\langle s_{v}|H_{0}|s_{v}\rangle=\sum_{\sigma'>\sigma}\,
\frac{\mathcal{E}
(\sigma,\sigma')}{\sigma' -\sigma}
\,\int d{\bf q}\,\left({\bf q}^{2}+ \frac{1}{2}
 m^{2}\right) A^{2}({\bf q})\,\int d{\bf q'}\,A^{2}({\bf q'}),
\ee
and,
\be
\langle s_{v}|H_{I}|s_{v}\rangle= 2 g\,B\,\sum_{\sigma} \mathcal{V}(\sigma)
\, \int d{\bf q}\, A({\bf q}).
\ee
We note that in deriving eq.(31), one has to expand the expression
$({\bf q}- {\bf q'})^{2}$ in eq.(15). 
 The cross term ${\bf q}\cdot {\bf q'}$ in this expansion does not
contribute because of the rotation invariance of the $A$'s.

The indicated sums over $\sigma$ and $\sigma'$ are easily done, after
$\rho(\sigma)$ in eqs. (14) and (21) is replaced by $\rho_{0}$, a $\sigma$
independent constant. We have,
\be
\sum_{\sigma'>\sigma}\,
\frac{\mathcal{E}(\sigma,\sigma')}{\sigma' -\sigma}
=\sum_{\sigma} 
\sum_{n=0}^{\infty}\frac{(1-\rho_{0})^{n}}{a (n+1)}=
 - \frac{p^{+}}{a}\,\frac{\ln(\rho_{0})}
{a\,(1-\rho_{0})},
\ee
and therefore,
\be
\langle s_{v}|H_{0}|s_{v}\rangle = \frac{p^{+}}{a}\left(-\frac{
\rho_{0}\,\ln(\rho_{0})}
{2 a (1-\rho_{0})}\,\int d{\bf q}\,(2 {\bf q}^{2}+ 
 m^{2}) A^{2}({\bf q})\right),
\ee
where eq.(28) was used.

After setting $\rho=\rho_{0}$, $\mathcal{V}$ becomes $\sigma$ independent,
but  it still depends on $\rho_{0}$.
We are unable to express this dependence
 in terms of elementary functions, so instead, we define,
\be
W(\rho_{0})=\sum_{n_{1}=0}^{\infty} \sum_{n_{2}=0}^{\infty} \frac{
(1-\rho_{0})^{n_{1}+ n_{2}+ 1}}{\sqrt{(n_{1}+ 1)(n_{2}+ 1)(n_{1}+
n_{2}+2)}},
\ee
and write,
\be
\langle s_{v}|H_{I}|s_{v}\rangle =2\, \frac{p^{+}}{a}\, 
\frac{g}{\sqrt{8 a^{3}}} \,\sqrt{1- \rho_{0}}\, W(\rho_{0})\,
\int d{\bf q}\, A({\bf q}).
\ee

In deriving eqs.(33) and (35), we have assumed that the upper limit of the
sum over the $\sigma$'s can be extended to to infinity, whereas in
reality, there is a cutoff at $\sigma=p^{+}$. For example, $W$ has
a spurious singularity as $\rho_{0}\rightarrow 0$,
\be
W(\rho_{0})\rightarrow \pi^{3/2}\,(\rho_{0})^{-1/2},
\ee
which would be absent if the cutoff is imposed.
Thus, the neglect of this
cutoff makes a difference only for small values of $\rho_{0}$.
Although it is not difficult to take care of the cutoff, the resulting
formulas become complicated. In the interests of simplicity, we will work
with the expressions derived above, keeping in mind the restriction that
they are not reliable for small values of $\rho_{0}$.

Putting together eqs.(30),(34) and (36), $I$ given by
\bq
I&=& -\frac{p^{+}\,\ln(\rho_{0})}{2 a^{2} (1 -\rho_{0})}\,
\int d{\bf q}\,(2 {\bf q}^{2}
+ m_{0}^{2})\,A^{2}({\bf q})\nonumber\\
&+&\frac{2 p^{+}\,g_{0}}{\sqrt{8 a^{5}}}\,\sqrt{1 -\rho_{0}}\,
W(\rho_{0})\,\int d{\bf q}\,A({\bf q})
+ \lambda\,\left(\int d{\bf q}\, A^{2}({\bf q}) -\rho_{0}\right).
\eq
 The first variational equation,
\be
\frac{\delta I}{\delta A({\bf q})}=0,
\ee
determines the functional dependence of $A$ on ${\bf q}$:
\be
A({\bf q})=- \frac{g_{0}}{\sqrt{2 a}}\,\frac{\sqrt{1- \rho_{0}}\,
W(\rho_{0})}{\tilde{\lambda}+ F(\rho_{0})(2 {\bf q}^{2}+ m_{0}^{2})},
\ee
where we have defined,
\be
\tilde{\lambda}=\frac{2 \lambda\, a^{2}}{p^{+}},\,\,\,
F(\rho_{0})= -\frac{\rho_{0}\,\ln(\rho_{0})}{1- \rho_{0}}.
\ee
In anticipation of renormalization, when one has to distinguish
between bare and renormalized constants,  $m$  and $g$ were replaced by
$m_{0}$ and  $g_{0}$.

From the above equation, it is easy to see that $A$ has the form of
a free light cone propagator. In fact, setting,
$$
A({\bf q})= \frac{const.}{{\bf q}^{2}+ m_{r}^{2}},
$$
where,
\be
m_{r}^{2}=\frac{m_{0}^{2}}{2}+\frac{\tilde{\lambda}}{2 F(\rho_{0})},
\ee
we can  identify $m_{r}$ with the renormalized mass.

Next, varying $I$ with respect to $\lambda$, reproduces the
normalization condition (28). We will see that the resulting
equation fixes the value of $\rho_{0}$, the parameter of interest.
Replacing $A$ by  (40), the integral involving
$A^{2}$ can be explicitly evaluated. At this point,
 the dimension of the transverse
space, which has been so far left arbitrary, has to be specified. The
critical dimension of the $\phi^{3}$ theory is six, corresponding to
$D=4$, accordingly,
in what follows, we will take $D=4$, although we could also have  studied
lower dimensions. The integral in question is then logarithmically
divergent, and it will be regulated by
cutting it off at $|{\bf q}|=\Lambda$. From now on, we will focus on the
cutoff dependent terms in the equations, and generally neglecting the
finite corrections. We will see that the finite terms can be absorbed
into the definition of the renormalized parameters.

 Evaluating the  logarithmic term in the integral
\be
\int d^{4} q A^{2}({\bf q})\rightarrow 2 \pi^{2}\int_{0}^{\Lambda} dq\,
q^{3} A^{2}({\bf q}) \approx \frac{g_{0}^{2}\,\pi^{2}\,(1-\rho_{0})\,
W^{2}(\rho_{0})}{4\,a\,F^{2}(\rho_{0})}\,\ln\left(\frac{\Lambda}
{\mu}\right),
\ee
the normalization condition reads,
\be
\rho_{0}=  \frac{g_{0}^{2}\,\pi^{2}\,(1-\rho_{0})\,
W^{2}(\rho_{0})}{4\,a\,F^{2}(\rho_{0})}\,\ln\left(\frac{\Lambda}
{\mu}\right)+ finite\, terms.
\ee

Here $\mu$ is mass parameter whose value depends on the finite terms
discussed above. Although these terms can be evaluated,  in the
 subsequent development, we will mainly be interested
in the cutoff dependence and renormalization. In fact,
if so desired, one can completely absorb the finite terms into
the definition of $\mu$.

From its definition as a probability, we know that $\rho_{0}$ lies
between zero and one and therefore cannot depend on the cutoff
$\Lambda$. This can be arranged by defining the renormalized coupling
constant by
\be
g_{r}^{2}= g_{0}^{2}\,\ln\left(\frac{\Lambda}
{\mu}\right).
\ee
This equation is consistent with the one loop renormalization group
equation, leading to asymptotic freedom, and in agreement with the
fact that $\phi^{3}$ theory is known to be asymptotically free in six
dimensions.

At this point, a natural question arises: How is it that in a phase
dominated by graphs of infinite order, the ultraviolet divergences agree
 with those deduced from perturbation theory? This is somewhat of a surprize,
and at least a partial explanation is the following. The calculation
presented here amounts to expanding the  fields $\phi$ and $\phi^{\dagger}$
about the classical background $A({\bf q})$, given by eq.(40). In this
sense, it is very similar to the usual field theory calculations in the
presence of a classical background, such as expanding about an instanton
in non-abelian gauge theories. In these calculations with a classical
background, the ultraviolet divergences are the same as in perturbation
theory, and the standard explanation is that the ``soft'' background
does not change the leading short distance singularities. Presently, work
is in progress to reformulate the variational calculation as a
systematic expansion around the classical background. Such a
reformulation, if successful, would substantiate the explanation
presented above.

Eq.(44) can be rewritten in terms of the renormalized coupling constant as
\be
\rho_{0}=\frac{\tilde{g}^{2}\,(1 -\rho_{0})\,W^{2}(\rho_{0})}
{F^{2}(\rho_{0})},
\ee
where, for convenience, we have defined,
$$
\tilde{g}^{2}=\frac{g_{r}^{2}\,\pi^{2}}{4\,a}.
$$

This is the equation that determines $\rho_{0}$. The question is then
whether this equation has a solution for $0\leq \rho_{0}\leq 1$. To
further simplify matters, we take the square root of (46) and define,
$$
\rho_{0}=x, \,\,\,x\,F(x)=L(x),\,\,\, \sqrt{x(1-x)}\,W(x)= Z(x),
$$
and rewrite it as,
\be
L(x)=|\tilde{g}|\,Z(x).
\ee

As shown in Fig.6, $L(x)$ is zero at $x=0$, and steadily increases to
$L=1$ at $x=1$.
\begin{figure}[t]
\centerline{\epsfig{file=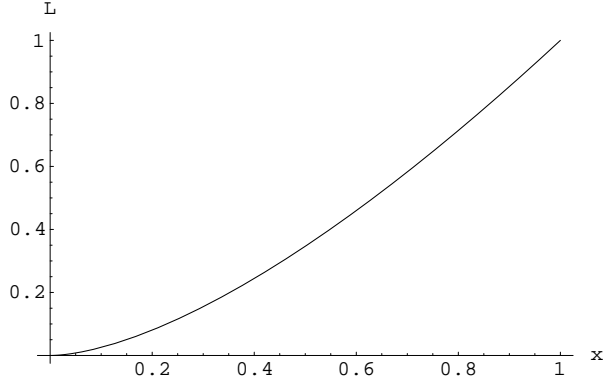,width=8cm}}
\caption{The Function L(x)}
\end{figure}
 On the other hand, $Z$ is equal to $\pi^{3/2}$ at $x=0$
(see eq.(37)),
and steadily goes down to $Z=0$ at $x=1$. $Z$ is plotted for $x>0.1$ in
Fig.7.
\begin{figure}[t]
\centerline{\epsfig{file=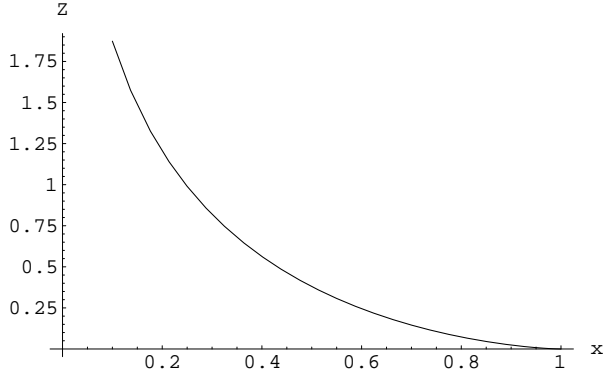,width=8cm}}
\caption{The Function Z(x)}
\end{figure}
 Clearly, if $\tilde{g}\neq 0$,
the two curves must intersect for some $0<x<1$, as shown in Fig.8 for
$\tilde{g}=1$.
\begin{figure}[t]
\centerline{\epsfig{file=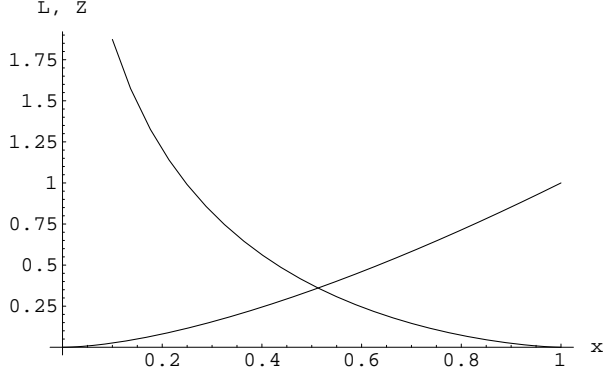,width=8cm}}
\caption{The Combined Plot Of L(x) And Z(x)}
\end{figure}
 This shows that $\rho_{0}=x\neq 0$, and therefore
 a condensate forms for non-zero
renormalized coupling. As $\tilde{g}\rightarrow 0$, the intersection point
approaches $x=0$, and explained in the paragraph after eq.(36),
 the expressions for $L$ and $Z$ become unreliable. In any case, as the
coupling tends to zero, we expect perturbation theory to take over
and the condensate to dissolve.

The third and the final variational equation, combined with eq.(42),
enables one to express the bare mass and the parameter
$\tilde{\lambda}$ in terms of $m_{r}$ and $g_{r}$.
It is gotten by varying $I$ with respect to $\rho_{0}$:
\bq
\frac{2 a^{2}}{p^{+}}\,\frac{\partial I}{\partial \rho_{0}}
&=&\sqrt{\frac{2}{a}}\,g_{0}\,
\left(\sqrt{1 -\rho_{0}}\,W(\rho_{0})\right)'\,\int d{\bf q}\,
A({\bf q})\nonumber\\
&+&F'(\rho_{0})\,\int d{\bf q}\,
\left(2 {\bf q}^{2}+ m_{0}^{2}\right)\,A^{2}({\bf q})
 -\tilde{\lambda}=0,
\eq
where the prime indicates the derivative with respect to $\rho_{0}$.
Substituting eq.(40) for $A$, this time we encounter quadratic divergences
at $D=4$. Introducing the cutoff $\Lambda$
 as in eq.(43), and keeping only the leading
quadratic divergence, gives the result,
\be
\rho_{0}\,F'\,m_{0}^{2} -\tilde{\lambda}= \frac{g_{0}^{2}\,
\pi^{2}\,\Lambda^{2}}{4 a \,F^{2}}\,\left(F\,\left((1 -\rho_{0})\,W^{2}
\right)' - (1-\rho_{0})\,F'\,W^{2}\right)+ finite\, terms.
\ee

 One can now solve the two linear equations (42) and (49)
 for $m_{0}^{2}$ and $\tilde{\lambda}$, and express them in terms of
$g_{r}$ (replacing $g_{0}$ by $g_{r}$ through (45)), $m_{r}$ and the cutoff
$\Lambda$. The bare mass and coupling are thus expressed in terms of the
renormalized mass and coupling and the cutoff; therefore, at least in this
approximation, the model is successfully renormalized.

Part of the the calculation presented above has obvious overlap with
perturbation theory: For example, the identification of $A({\bf q})$
with the free propagator, the cutoff dependence of bare parameters and
the relation between bare and renormalized couplings (asymptotic freedom)
all parallel standard perturbative results. On the other hand,
our variational calculation goes beyond perturbation theory. For
example, the functions
$F$ and $W$ have no obvious perturbative interpretation. In particular,
we have argued earlier that $\rho_{0}$ is zero in any finite order
of perturbation, whereas the eq.(46) has a solution $\rho_{0}\neq 0$ for
$g_{r}\neq 0$. The formation of a condensate is therefore clearly
a non-perturbative phenomenon.

A somewhat surprising feature of the variational ansatz (27) is that
it is not an eigenstate of the total momentum operator
\be
{\bf P}=\sum_{\sigma}\int d{\bf q}\,
{\bf q}\,\phi^{\dagger}(\sigma,{\bf q})\,
\phi(\sigma,{\bf q}).
\ee
Only the expectation value of ${\bf P}$ is well defined and it is zero:
\be
\langle s_{v}|{\bf P}|s_{v} \rangle =0.
\ee
This follows from the rotation invariance of the function
$A({\bf q})$. Of course, this does not mean that conservation of
momentum is violated. The real ground state is an eigenstate of momentum
with eigenvalue zero. What 
 this really means is that the trial ground state we are using
is a superposition of states with different momenta, with only the 
average momentum fixed at zero. States of this type are commonly used
in statistical mechanics to construct the grand canonical ensemble;
an ensemble of states with varying particle number is a typical example.
We note that, since the boundary conditions at the initial and final
times  were left free, world sheet configurations of arbitrary total
momenta are allowed. Of course,
we could constrain the total momentum of the trial state to be zero by,
for example, setting the momentum at a particular site to be minus the
sum of the momenta at the other sites. However, this would complicate
the variational calculation considerably, and in the limit of large
number of sites, it would probably not change the final outcome significantly.
In any case, whether a trial wave function approximates the real
one reasonably well is always an open question.

Finally, a few words about the decompactification limit $a\rightarrow 0$
are in order. At the beginning, $N_{0}$ (eq.(25)) was specified to be
large but finite. It had to be large to be able to define a condensate
in the first place, and it had to be finite to avoid possible infrared
singularities. The question arises: Are there singularities in the limit
$a\rightarrow 0$ ? Consider the equations we have derived in this section,
for example, equations (46) and (49). Expressed in terms $\tilde{g}$, these
equations do not depend on $a$ anymore. So a simple redefinition of the
coupling constant eliminates possible singulaties. Of course, this is not the
whole story: Higher order corrections to our approximation may turn
to be singular in the decompactified model. We hope to study this problem
in the future.

\vskip 9pt

\noindent{\bf 6. Conclusions}

\vskip 9pt

The present work is a fresh approach to the old problem of the summation
of planar graphs [9,10]. Although it has some overlap with
 the earlier work by Thorn and the present author [1,2,3,4], especially with
reference [4], it is essentially a fresh start. The main difference with
the earlier work is that it is based on second quantization rather than
first quantization. Just as in the usual second quantized field theory,
the Hamiltonian is expressed in terms of creation and annihilation
operators, except that these operators, instead of particles, create and
annihilate boundaries on the world sheet. With the help of a simple
variational ansatz, we were able to investigate the structure of the
 ground state. In particular, we were able to show that the ground
state was a condensate of boundaries (solid lines). This calculation
was carried out in six dimensional space-time, which is the critical
dimension of the $\phi^{3}$ theory. The ultraviolet divergences
encountered in the course of the variational calculation are
completely consistent with standard perturbation results, and they
can be renormalized in the usual fashion.

We hope to follow up on the present work along various directions.
For example, either using a variational or a mean field method, it
should be possible to go beyond the ground state of the model
and investigate the whole spectrum of excited states. This should give us 
information about possible string formation, which we have not
discussed here. Another interesting future project is to study the limit
of continuum world sheet by decompactifying the model. In this limit,
one expects a simpler local field theory to emerge from the
non-local model presented here.
Finally, building on the earlier work [11,12], we hope to incorporate
a physical theory such as QCD into the framework  developed here.

\vskip 9pt

\noindent{\bf Acknowledgement}

\vskip 9pt

This work was supported by the Director, Office of Science,
 Office of High Energy  Physics, 
 of the U.S. Department of Energy under Contract DE-AC02-05CH11231.

\vskip 9pt

{\bf References}

\vskip 9pt

\begin{enumerate}
\item K.Bardakci and C.B.Thorn, Nucl.Phys. {\bf B 626} (2002) 287,
hep-th/0110301.
\item K.Bardakci and C.B.Thorn, Nucl.Phys. {\bf B 652} (2003) 196,
hep-th/0206205.
\item K.Bardakci, Nucl.Phys. {\bf B 715} (2005) 141, hep-th/0501107.
\item K.Bardakci, JHEP {\bf 0807} (2008) 057, arXiv:0804.1329.
\item G.'t Hooft, Nucl.Phys. {\bf B 72} (1974) 461.
\item K.Bardakci, Nucl.Phys. {\bf B 667} (2004) 354, hep-th/0308197.
\item T.Banks, W.Fischler, S.H.Shenker and L.Susskind, Phys.Rev.
{\bf D 55} (1997) 5112, hep-th/9610043.
\item L.Susskind, hep-th/9704080.
\item H.B.Nielsen and P.Olesen, Phys.Lett. {\bf B 32} (1970) 203.
\item B.Sakita and M.A.Virasoro, Phys.Rev.Lett. {\bf 24} (1970) 1146.
\item S.Gudmundsson, C.B.Thorn and T.A.Tran, Nucl.Phys. {\bf B 649}
(2003) 3, hep-th/0209102.
\item C.B.Thorn and T.A.Tran, Nucl.Phys. {\bf B 677} (2004) 289,
hep-th/0307203.
\end{enumerate}

\end{document}